# Rapidly Achieving Chemical Accuracy with Quantum Computing Enforced Language Model

Honghui Shang[1†], Xiongzhi Zeng[2†], Ming Gong[2,3,4†], Yangju Wu[1],
Shaojun Guo[2,3,4], Haoran Qian[2,3,4], Chen Zha[2,3,4], Zhijie Fan[2,3,4], Kai Yan[2,3,4],
Xiaobo Zhu[2,3,4], Zhenyu Li[1,4*], Yi Luo[2,4*], Jian-Wei Pan[2,3,4*] and Jinlong Yang[1,4*]

[1]Key Laboratory of Precision and Intelligent Chemistry,
University of Science and Technology of China, Hefei 230026, China

[2]Hefei National Research Center for Physical Sciences
at the Microscale and School of Physical Sciences,
University of Science and Technology of China, Hefei 230026, China

[3]Shanghai Research Center for Quantum Science and CAS Center
for Excellence in Quantum Information and Quantum Physics,
University of Science and Technology of China, Shanghai 201315, China

[4]Hefei National Laboratory,
University of Science and Technology of China, Hefei 230088, China

[†]These authors contribute to this work.
*To whom correspondence should be addressed:
zyli@ustc.edu.cn; yiluo@ustc.edu.cn;
pan@ustc.edu.cn; jlyang@ustc.edu.cn

**Finding accurate ground state energy of a many-body system has been a major challenge in quantum chemistry. The integration of classic and quantum computers has shed new light on resolving this outstanding problem. Here we propose QiankunNet-VQE, a transformer based language models enforced with quantum computing to learn and generate quantum states. It has been implemented using up to 12 qubits and attaining an accuracy level competitive with state-of-the-art classical methods. By leveraging both quantum and**

**classical resources, this scheme overcomes the limitations of variational quantum eigensolver (VQE) without the need for cumbersome error mitigation. Moreover, QiankunNet-VQE provides a different route to achieve a practical quantum advantage for solving many-electron Schrödinger equation without requiring extremely precise preparation and measurement of the ground-state wavefunction on quantum computer.**

## One-Sentence Summary

We combine the quantum computing with QiankunNet to provide a scalable way to attain accurate ground state energy with limited quantum resources.

## Introduction

The electronic structure and properties of quantum materials can, in principle, be determined by solving the many-electron Schrödinger equation. The most challenging task is finding a general approach to reduce the exponential complexity of the full many-body wave function and extract its essential features. Various methods have been developed to solve the Schrödinger equation for realistic complex systems, including density functional theory (1), wavefunction-based methods like Hartree-Fock (HF) (2) and post-Hartree-Fock techniques such as configuration interaction (CI) (3,4), coupled cluster (CC) (5), density matrix renormalization groups (6) and quantum monte carlo (QMC) (7,8). However, the accuracy of these methods is often unsatisfactory and may even fail in numerous cases, mostly due to the limited expressive power of the wave function ansatz. Quantum computing exhibits tremendous potential in the field of quantum chemistry, exemplified by the quantum phase estimation algorithm (9,10), which theoretically has the capability to describe chemical systems with any desired precision, provided there are sufficient fault-tolerant qubits. On the other hand for solving quantum chemistry prob-lems on the noisy intermediate-scale quantum (NISQ) devices, the variational quantum eigensolver(VQE) is an appealing candidate (11), which has a great flexibility in choosing quantum circuit ansatzes and mitigating errors. Nonetheless, its current capabilities are constrained by the limited quantum resource, the optimization difficulties, measurement overhead and circuit noise under existing technology (12).

The most promising paths to practical quantum advantage in the NISQ era is by leveraging more efficient classical optimization method beyond VQE to overcome hardware limitations, Huggins et al. (13) have proposed a hybrid quantum-classical algorithm that combines a quantum trial wave function generated by Sycamore quantum processor with constrained quantum monte carlo approach to solve quantum chemistry problems. However, the accuracy of constrained QMC calculations is ultimately constrained by the quality of the trial wave function, and the use of arbitrary Slater determinants as the trial wave function could potentially be supplanted by unitary coupled-cluster ansätze to enhance the approach's accuracy. Zhang et al. (14) employ artificial neural networks to enhance the performance of VQE by enabling the efficient representation of quantum states. They jointly optimize the VQE parameters and the neural network parameters to improve the convergence of quantum computations. However, this approach requires measurements of the quantum state to be performed in each iteration, leading to a high experimental overhead.

Recently, the Neural Network Quantum States (NNQS) methods have proven to be powerful tools in the realm of quantum chemistry. These methods efficiently capture complex relationships for quantum systems within exponentially large encoded Hilbert spaces (15), leading to their widespread adoption in tasks such as molecular energy prediction, electronic structure calculations, and quantum wavefunction modeling (16–19). We have recently proposed Qiankun-Net (20), a transformer-based neural-network architecture along with batched autoregressive sampling, which could significantly improve the accuracy of first-principles calculations compared to existing fermionic ansatzes. Given the success of transformer-based language models in natural language processing, it is natural to adapt large language model architectures to tackle quantum physical problems. These architectures can provide expressive and flexible parameterizations of quantum states, which can be optimized using powerful heuristics. However, NNQS methods face the problem of slow convergence rates. Quantum computers can just provide an efficient initial quantum wave function, assisting NNQS in finding solutions to quantum systems more rapidly. Recently, Bennewitz et al. (21) have adopted near-term quantum simulations and NNQS to improve estimates of ground states energy of $H_2$ and LiH molecules, which only requires samples from measurements on the quantum state once, and then train the weights in the neural network to match the distribution produced via VQE as closely as possible, serving as an initial step before continuing the NNQS optimization process. However, their approach shares the same neural network architecture for both the wavefuction phase and amplitude, which increases the reinforcement learning complexity and still faces the problem of slow convergence. Moreover, they use an autoregressive sampling method in NNQS, which creates many repetitive samples and thus hinders the scalability of their approach.

In order to address the above challenges, we propose here a new scheme called QiankunNet-VQE, which combines quantum computing with a transformer-based language model. QiankunNet-VQE uses quantum computers to make the kernel part of system to become entanglement states, and then the transformer neural networks are used to perform the training and inference for the whole system. QiankunNet-VQE have several advantages: 1. QiankunNet-VQE requires a series of low-depth quantum ansatz VQE experiments in a noisy environment, focused within the active space to approximate the ground state closely. This approach conserves quantum resources, allowing the extraction of desired states from the resulting variational parameters and circuits without the need to repeat the VQE experiments. 2. QiankunNet-VQE efficiently tackles the issue of low measurement precision encountered during the estimation of quantum observable with NISQ devices, and thanks to VQE, QiankunNet-VQE can use systematically improvable trial wavefunctions, such as unitary coupled-cluster state by incorporating of additional excitation operators. 3. QiankunNet-VQE can rapidly converge to the true ground state of a chemical system. QiankunNet-VQE uses key quantum states obtained from VQE simulation or experiments, and compared with using random initial guesses, the convergence speed can be improved by almost an order of magnitude. By employing a generative transformer architecture, we not only improve expressiveness but also leverage its autoregressive nature to replace the less efficient Markov Chain Monte Carlo (MCMC) sampling with a more efficient batched autoregressive algorithm. 4. QiankunNet-VQE can accurately calculate electron correlation energies in the full Hilbert spaces that are significantly larger than what the number of qubits used in VQE, which can handle large number of orbitals while constraining the quantum resources to a much smaller active space without incurring additional measurement overheads. It not only addresses current challenges in quantum chemistry but also opens new avenues for exploration and discovery in the NISQ era and beyond.

## Theory and algorithms: the design of QiankunNet-VQE

The central task for many-body quantum physics/chemistry calculation is to solve the static Schrödinger equation $\hat{H}|\Psi\rangle = E|\Psi\rangle$ to get the ground state $|\Psi\rangle$ and the ground-state energy $E$ of the many-body interacting Hamiltonian:

$$\hat{H} = -\sum_{i=1}^{N} \frac{1}{2}\nabla_i^2 - \sum_{i=1}^{N}\sum_{A=1}^{M} \frac{Z_A}{|\mathbf{r}_i - \mathbf{R}_A|} + \sum_{i=1}^{N}\sum_{j>i}^{N} \frac{1}{|\mathbf{r}_i - \mathbf{r}_j|} \tag{1}$$

where $N$, $M$ denote the total number of electrons and nuclei, $\nabla_i$ is the single particle kinetic operator of the $i$-th electron, $\mathbf{r}_i$ indicates the electronic coordinates, $\mathbf{R}_A$ and $Z_A$ indicate the coordinates and charges of the $A$-th nucleus in the molecule.

The many-electron Schrödinger equation can be solved in quantum computer with VQE, which begin with a parameterized wavefunction on a quantum computer, often expressed as $|\Psi\rangle = U(\Theta)|\Psi_0\rangle = \prod_{i=1}^{N} U_i(\theta_i)|\Psi_0\rangle$, $U_i(\theta_i)$ is the parameterized quantum circuits or quantum machine learning circuit with L layers which can either be associated with a specific electron excitation or correspond to a set of quantum gates, depending on whether a physically motivated unitary coupled-cluster or a hardware heuristic ansatz is adopted (22). $|\Psi_0\rangle$ is a reference state such as the Hartree-Fock state. In both cases, the ground-state energy can be estimated by variationally minimizing the expected value of the Hamiltonian. The practicability of current various NISQ devices is limited by many factors like the short coherent time, number of qubits and external noise and so on (23,24). These adverse factors will cause the calculation results to deviate from expectations and cause huge deviation (25). Another disadvantage is the number of Hamiltonian terms that grow with the polynomial number of electrons. In the process of VQE experiment, the accumulation of these terms will lead to the accumulation of noise, which will cause a large deviation in the results. From analysis of quantum chemistry, the ground state Hamiltonian has many terms with very small coefficients. While the individual coefficients of these terms are small, resulting in a relatively minor impact on the energy, their sheer quantity, when subjected to the accumulation of noise, can lead to substantial deviations from the actual energy.

Our QiankunNet-VQE method, illustrated in Fig. 1, is composed of two steps. First, we start from the active space method and choose the correlation orbitals, and then tailor the Hamiltonian to remove these small terms for VQE experiment. After the active space selection and tailoring of Hamiltonian, a VQE is performed with a well-designed short circuit ansatz under with noise devise (25). We can extract essential configuration information from the results through selective sampling on a classical computer. Using the resultant variational parameters from the VQE experiment, we can reconstruct the important wavefunction vectors from the quantum circuit.

The choice of active orbitals plays a crucial role in determining the accuracy and computational feasibility of the calculation, taking the VQE simulation of LiH molecule as an example, using a minimal basis set, the LiH molecule has a total of six molecular orbitals, requiring twelve qubits for a complete representation. However, some molecular orbitals may contain very deep energy levels and their contributions to the total correlation energy are less significant compared to those of the molecular orbitals related to chemical reactions processes (active space orbitals). And the majority of the configuration coefficients are concentrated on configurations dominated by active space orbitals. Consequently, in VQE calculations, it is common practice to select a subset of active orbitals to reduce the computational complexity while still capturing the essential chemical properties of the molecule. In this specific example, we consider a scenario where only three active orbitals (six qubits) are selected for the VQE experiment. The accuracy of the VQE results in this reduced active space can be assessed by comparing the lowest eigenvalue obtained through exact diagonalization within the active space to the true Full Configuration Interaction (FCI) value in the complete space. Our calculations reveal a noticeable gap between these two values, amounting to around 4/12 millihartrees (mHa) for the LiH/$F_2$ molecule, as shown in Table S5/S12. The discrepancy between the lowest eigenvalue in the active space and the true FCI value in the complete space can hinder the ability of VQE experiments to achieve chemical accuracy, even when error-mitigation techniques are employed (25).

Subsequently, such selected configurations and coefficients prepared by a noisy quantum device are used to pre-training the transformer-based neural network (QiankunNet), which is an autoregressive generative neural network. QiankunNet has the capacity to learn molecular wavefunctions by representing the quantum state of a molecule as a sequence of tokens, is trained using a blend of supervised techniques to learn the quantum state knowledge of quantum computer. After the pre-training process, we further use the variational Monte Carlo algorithm with the same QiankunNet ansatz to improve the representation of the many-body quantum ground state. QiankunNet adopts the attention mechanism within the Transformer architecture and has the capability to model long-range temporal and spatial correlations, this architecture has been used in many state-of-the-art experiments in natural language and image processing and has the potential to model long-range quantum correlations. During this optimization process, QiankunNet is presented with a sequence of molecular orbital embeddings, which encode the quantum state of each atom orbital in the molecule, and self-supervise learning is adopted to create correct quantum state. This capability facilitates accurate predictions of electronic properties and lays the foundation for further quantum chemical applications. Our QiankunNet-VQE algorithm not only harnesses the power of quantum computers in their current imperfect state but also accelerates the performance of neural networks.

For further details, we refer the reader to the Methods and Supplementary Information for a complete description of VQE, VMC and the Transformer neural network (QiankunNet). Numerous numerical and experimental results demonstrate the efficacy of our method.

## Results

In order to inspect the power of our QiankunNet-VQE, we start from LiH system to check the feasibility and advantage of QiankunNet-VQE over VQE experiment and NNQS method QiankunNet. Then, we demonstrate the application of QiankunNet-VQE to estimate the potential surface of $F_2$ system, then we investigate $H_4$ and $H_{10}$ with strong correlation.

**QiankunNet-VQE with quantum computing experiments for LiH**  As the first example, we demonstrate the efficiency of QiankunNet-VQE on LiH molecule involving 12 qubits, the quantum trial wavefunction is adopted from a UCC quantum circuit in the 6 qubits VQE experiment (25), we then acquired VQE experimental results, encompassing variational parameters of the quantum circuit (25). We perform the calculation using STO-3G basis set. Because of the influence of noise, the VQE experiments for large bond lengths fail to provide accurate results for the potential energy surface of the LiH system, even with the implementation of efficient error-mitigation methods (25). Nonetheless, the resulting quantum circuit yields essential configuration information, albeit without converged energy, see Supplementary Materials. We observe that, despite the presence of significant noise in VQE experiment, QiankunNet-VQE finally achieves chemical accuracy by improving the representation of the ground state on quantum computer. Our algorithms based on the VQE circuit deliver highly accurate results, with a faster convergence speed than QiankunNet, as illustrated in Fig. 2. To unravel the QiankunNet-VQE results on LiH further, in Fig. 2(b), we illustrate the probability distribution of the most relevant configurations of LiH. We contrast between the QiankunNet method and QiankunNet-VQE. It is clear from the histogram that the QiankunNet-VQE is able to predict more accurate correlations than QiangkunNet when comparing with FCI results. These outcomes underscore the feasibility and efficiency of our QiankunNet-VQE approach.

**$F_2$: recover accurate PES from VQE experiment** Due to the inherent complexity of chemical systems and the imperfections in NISQ experimental setups, the maximum number of qubits for VQE simulation is limited to 12 (23,25). An illustrative example of the capabilities and limitations of current NISQ computers is presented in the VQE experiment of $F_2$ as shown in Fig. 3. This system, featuring ten orbitals and eighteen electrons in the minimal basis set, can be simulated with 12 qubits by freezing the lowest four orbitals. However, the performance of the VQE algorithm is significantly affected by the relatively deep quantum circuit with numerous two-qubit gates. Despite employing various noise-mitigation techniques, we are unable to recover the accurate Potential Energy Surface (PES) from the quantum experiment (25). As shown in Fig. 3, QiankunNet-VQE improves the ground-state estimation of $F_2$ to chemical accuracy for all bond lengths, with the quantum circuit parameters from noisy VQE experiment.

**QiankunNet-VQE with quantum computing simulator for strong correlation systems: $H_4$ and $H_{10}$** We further examine QiankunNet-VQE for $H_4$ (8 qubits) and $H_{10}$ (20 qubits) systems with strong correlation. For $H_4$ system, we consider the square geometry with D4h symmetry, in which, two quasidegenerate state can be found, and poses a great challenge for single-reference methods, so it is a famous benchmark test for electron correlation method. We use the minimal (STO-3G) basis set, and perform calculations for various method. Through our computational experiment, single-reference methods such as unrestricted Hartree-Fock (UHF), configuration interaction with single and double substitutions (CISD), and unitary coupled cluster with single and double substitutions parametrized as a variational quantum circuit (UCCSD-VQE) fail to achieve chemical accuracy, even when combined with QiankunNet searches initialized from UHF or CISD states. Randomly initialized QiankunNet is capable of achieving chemical accuracy, although more slowly than a hybrid QiankunNet-VQE approach. Remarkably, by combining variational quantum eigensolver with neural network quantum states, chemical accuracy can be reached more rapidly during the optimization process. Specifically, The ground-state wave function of strong correlation system is related to many configurations with larger configuration coefficients while it is only related to a few key configurations for weakly correlated system. These representative single-reference methods like UHF,CISD,UCCSD, all have inherent limitations, namely, they cannot simultaneously consider the contributions of multiple configurations in finite truncation cases, thus failing to achieve high accuracy. Using random initialization has a certain probability of obtaining the key configurations, and QiankunNet can achieve high accuracy through sampling, but the convergence speed is slow. VQE can advance the provision of these configurations and their non-zero configuration coefficients, and by pre-training, QiankunNet can accelerate convergence. This hybrid method thereby provides an advantageous route towards rapid determination of strongly correlated molecular ground states to chemical precision on near-term quantum devices.

We further examine a larger strongly correlated system, specifically the potential energy surface of the $H_{10}$ chain system, which is a spaced linear hydrogen chain, and is of great interest since it is a system which can bridges model Hamiltonians and real material systems. Fig. 4(b) presents a detailed comparison of potential energy curves (PEC) produced by various methods using a minimal STO-6G basis set (26). The PEC exhibits the typical features of initial short-range repulsion due to Coulomb repulsion and Pauli exclusion, followed by a minimum energy value reached at equilibrium bond length ($R_0$). Beyond $R_0$, the curves monotonically increase toward the asymptotic value of an isolated hydrogen atom's ground state energy, with well depth corresponding to dissociation energy. Owing to the small size and basis set, the potential curve could be obtained exactly using full configuration interaction (FCI). The lower panel of Fig. 4(b) provides a magnified view showing deviations from the FCI reference values. Because electronic correlations become more prominent as the inter-nuclear separation increases, thereby enhancing the multireference nature of the system's wavefunction in this region. Single-reference approaches such as UHF and UCISD, as well as UCC-VQE, are unable to adequately describe such static correlation effects across strong correlations regimes. While QiankunNet performs reasonably well at smaller R values, its accuracy deteriorates in the $2\,\text{Å} < R < 3\,\text{Å}$ interval where multireference character dominates. By combining the strengths of VQE and QiankunNet through a hybrid QiankunNet-VQE approach, chemical accuracy across the full potential energy surface can be achieved by effectively capturing the correlations inherent to this challenging multireference problem. These results can be directly compared with the most precise methods currently available, such as DMRG (Density Matrix Renormalization Group) and MRCI+Q (Multi-Reference Configuration Interaction with Quadruple excitations) (26), further demonstrating the suitability of our approach for strongly correlated systems.

## Discussions

Combining quantum computing with language model architecture represents an exciting frontier in the field of quantum computation and artificial intelligence. Quantum computing utilizes the principles of quantum mechanics to process information and perform computations in a fundamentally different way than classical computers. Neural network quantum states, on the other hand, leverage quantum systems to represent and process data for machine learning tasks. The idea behind combining quantum computing with quantum neural network states is to harness the unique properties of quantum systems to enhance the processing capabilities of neural networks. The proposed QiankunNet-VQE model represents the first scalable method to accelerate VQE with language model that is efficient to implement on NISQ hardware. A key challenge in quantum computing for quantum chemistry is achieving high-accuracy simulations with limited quantum resources. Traditional methods like VQE have been hampered by hardware constraints such as short coherence times. QiankunNet-VQE leverages both quantum and classical resources to overcome limitations of standard VQE and provide a scalable way to enhance the Ansatz without requiring additional quantum resources. Another critical aspect of QiankunNet-VQE is its scalability, enabled by its polynomial processing time overhead. This is particularly important considering the complexity and size of problems in quantum chemistry. The scalability of QiankunNet-VQE suggests that it can be adapted to more complex systems and larger molecular structures, which was a significant limitation with previous quantum computing methods. The success of QiankunNet-VQE can be attributed to its hybrid nature, which effectively leverages both quantum and classical resources. This synergistic approach is not just about augmenting quantum computations with classical algorithms; it's about creating a coherent system where each part complements and enhances the other. By doing so, QiankunNet-VQE overcomes many of the limitations currently faced in the NISQ era, such as error rates and hardware restrictions.

In summary, QiankunNet-VQE stands as a groundbreaking development in quantum machine learning. By adeptly combining the strengths of transformer neural networks with quantum computing, it not only addresses current challenges in quantum chemistry but also opens new avenues for exploration and discovery in the NISQ era and beyond. This hybrid model could well be the template upon which future quantum computing applications in various scientific fields are built. Looking forward, the exponential acceleration potential of QiankunNet-VQE, even with NISQ era hardware, is particularly promising. This suggests that as quantum hardware continues to evolve and improve, the capabilities of hybrid models like QiankunNet-VQE will similarly expand. The implications of this for quantum chemistry and other domains are profound, potentially leading to breakthroughs in material science, drug discovery, and complex molecular simulations.

# Acknowledgments

**Funding:** This work is supported by National Natural Science Foundation of China (Grant No. T2222026, 22303090, 22003073, 11805279, 22393913 and 21825302), by the Strategic Priority Research Program of the Chinese Academy of Sciences (XDB0450101). This work was supported by the Supercomputing Center of the USTC. **Data and materials availability:** The data that support the findings of this study are available from the corresponding author upon reasonable request.

# Supplementary Materials

Materials and Methods

Supplementary Text

Figure S1 to S14 ; Tables S1 to S15

References (1-26)

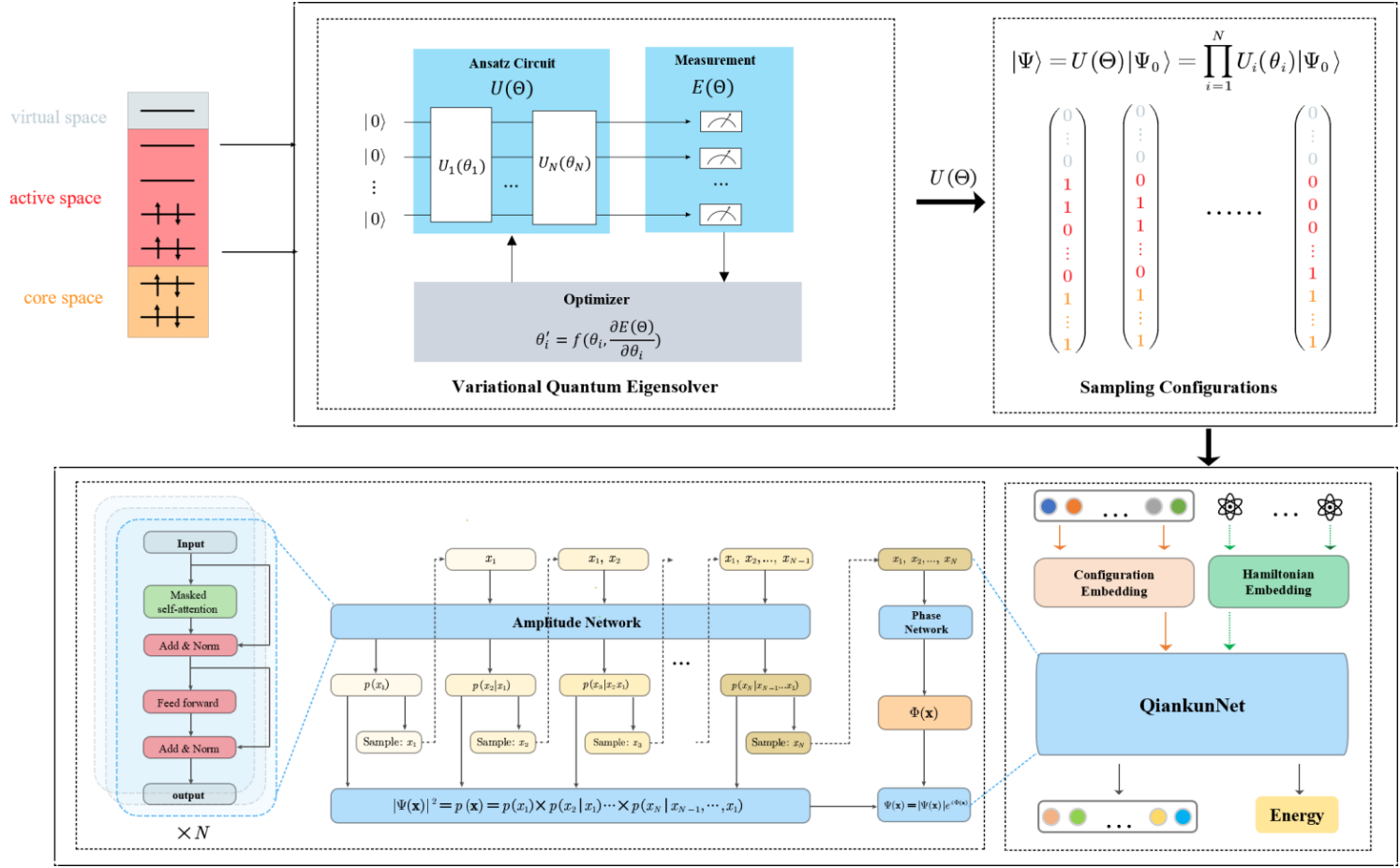

Figure 1: The flowchart of **QiankunNet-VQE**: First, an approximate ground state $|\Psi\rangle$ is prepared on a quantum computer by unitary evolution U of an initial quantum state $|\Psi_0\rangle$. Second, the parameters from VQE are used to train QiankunNet and the neural network ansatz is post-processed using QiankunNet to obtain the approximation of the ground-state energy of the Hamiltonian. The final ground state $|\Psi\rangle$ is returned at the end of the procedure. The Transformer architecture comprises multiple layers of self-attention and feedforward neural networks. The self-attention mechanism empowers the model to capture long-range interactions between atoms, while the feedforward networks introduce nonlinearity and depth to the model.

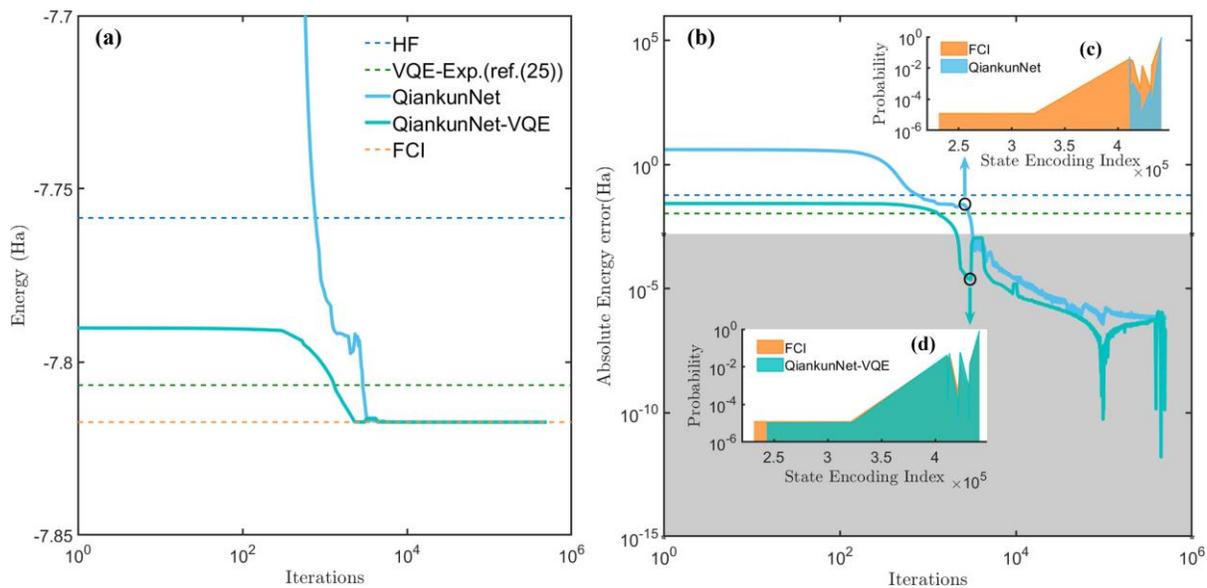

Figure 2: (a) The energy and (b) absolute energy error converged process of LiH with bond length of 2.6 Å for HF, VQE experiment i.e. Exp.(Ref. (25)), QiankunNet and QiankunNet-VQE.(c)(d)The probabilities (in logarithmic scale) of the configurations in the FCI (orange), QiankunNet (green) wavefunctions for the LiH molecule at bond length of 2.6 Å with QiankunNet and QiankunNet-VQE(blue).

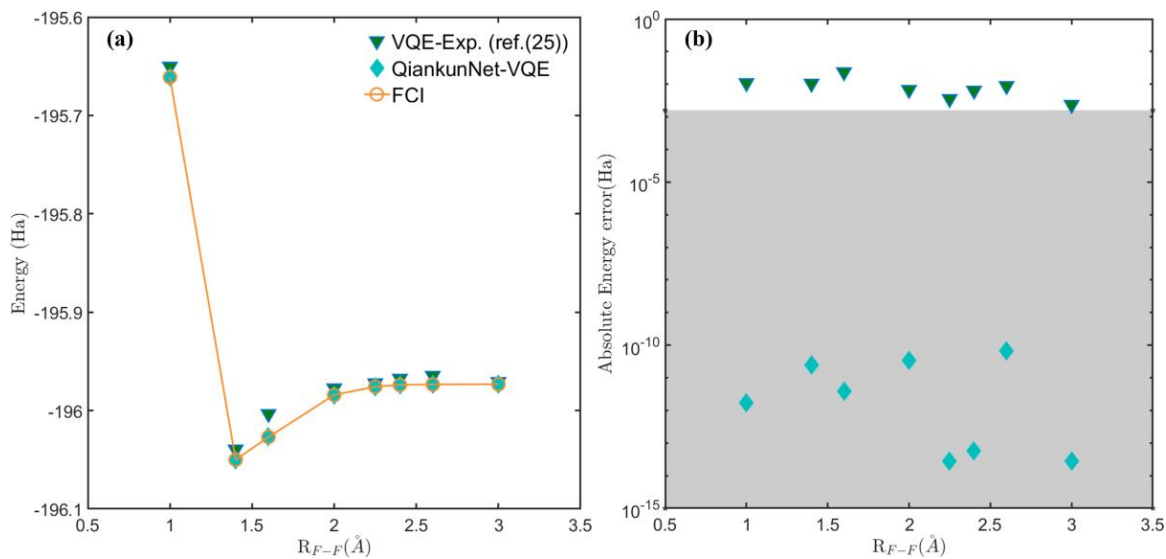

Figure 3: (a) The energy and (b) absolute energy error converged process of $F_2$ potential energy surface for VQE experiment i.e. Exp.(Ref. (25)), QiankunNet-VQE and FCI.

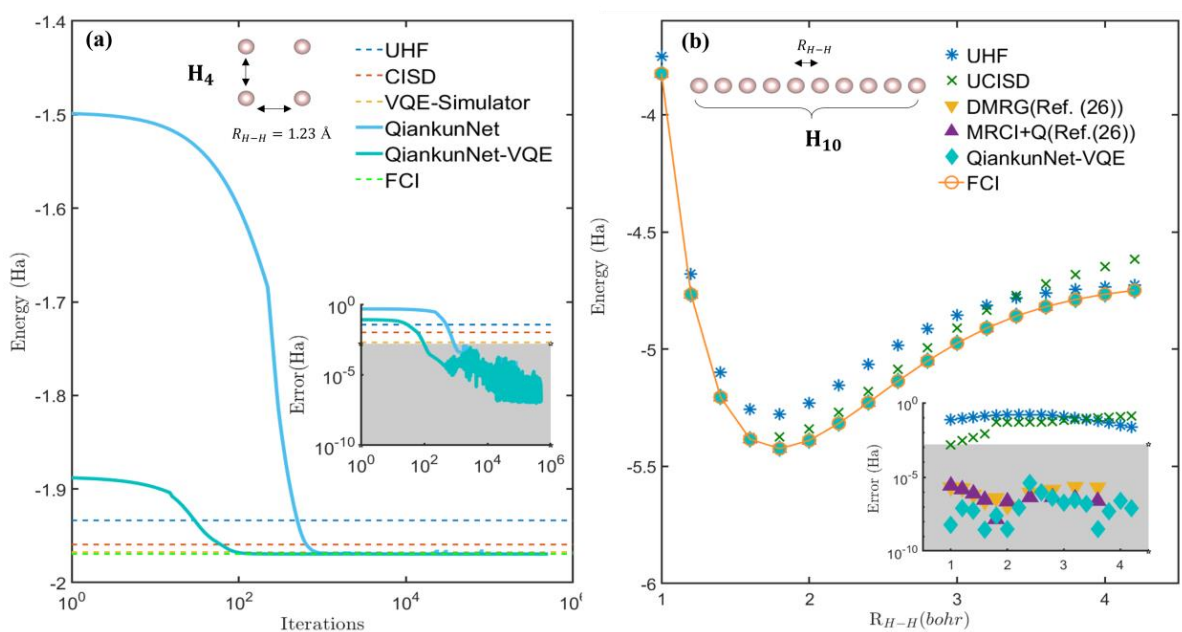

Figure 4: (a) The energy and absolute energy error (inner figure) converged process of $H_4$ molecule with bond length of 1.23 Å for UHF, UCISD, UCC-VQE, QiankunNet, QiankunNet-VQE and FCI. (b) Potential energy curve of $H_{10}$ molecule and deviations from FCI (inner figure) for UHF, UCISD, DMRG (26), MRCI+Q (26), QiankunNet-VQE and FCI.

# Supplementary Information for Rapidly Achieving Chemical Accuracy with Quantum Computing Enforced Language Model

Honghui Shang[1†], Xiongzhi Zeng[2†], Ming Gong[2,3,4†], Yangju Wu[1], Shaojun Guo[2,3,4], Haoran Qian[2,3,4], Chen Zha[2,3,4], Zhijie Fan[2,3,4], Kai Yan[2,3,4], Xiaobo Zhu[2,3,4], Zhenyu Li[1,4*], Yi Luo[2,4*], Jian-Wei Pan[2,3,4*] and Jinlong Yang[1,4*]

[1]Key Laboratory of Precision and Intelligent Chemistry,
University of Science and Technology of China, Hefei 230026, China
[2]Hefei National Research Center for Physical Sciences
at the Microscale and School of Physical Sciences,
University of Science and Technology of China, Hefei 230026, China
[3]Shanghai Research Center for Quantum Science and CAS Center
for Excellence in Quantum Information and Quantum Physics,
University of Science and Technology of China, Shanghai 201315, China
[4]Hefei National Laboratory,
University of Science and Technology of China, Hefei 230088, China

†These authors contribute equally to this work.
∗To whom correspondence should be addressed:
zyli@ustc.edu.cn; yiluo@ustc.edu.cn;
pan@ustc.edu.cn; jlyang@ustc.edu.cn

**Content:**

# 1. Technical Details and Algorithmic Flowchart

## 1.1 Challenges Faced by Classical Quantum Chemistry Methods

The exploration of high accuracy quantum chemistry methods has always been at the forefront of scientific research, offering profound insights into the molecular dynamics and properties of substances. However, classical quantum chemistry methods face significant challenges, particularly in terms of computational *scalability and accuracy* when dealing with complex systems. These traditional approaches often struggle to capture the intricate quantum mechanical interactions essential for accurate predictions at a larger scale (1,2). Among these methods, Hartree-Fock (HF), Configuration Interaction with Singles and Doubles (CISD), Coupled Cluster with Singles and Doubles (CCSD), Multi-Reference Configuration Interaction (MRCI) method and Full Configuration Interaction (FCI) are foundational, each with its unique approach to solving the electronic Schrödinger equation.

The HF method approximates the wave function of a many-electron system with a single Slater determinant. $\Psi_{HF}$ is expressed as:

$$|\Psi_{HF}\rangle = det|\phi_1 \phi_2 \ldots \phi_N|$$

where $\phi_i(j)$ represents the spin orbital of the ith electron. The energy is minimized according to the variational principle, leading to the Fock equations, which are solved iteratively. The HF energy is the sum of the one-electron energies and the electron-electron repulsion, minus the exchange energy. The computational complexity of HF is $O(N^4)$, where N is the number of basis functions. While HF provides a good starting point, its accuracy is limited because it does not account for electron correlation, often leading to significant errors in predicting molecular properties.

The CC method employs an exponential ansatz to incorporate correlation effects beyond HF in a systematically improvable manner. The CC wave function,

$$|\Psi_{CC}\rangle = e^T |\Psi_{HF}\rangle$$

T is the cluster operator that encompasses single ($T_1$), double ($T_2$), and potentially higher-order ($T_3, T_4, \ldots$) excitations. The exponential form of the operator ensures the inclusion of all possible excitation paths, providing a highly correlated description of the electronic structure. CCSD incorporates $T_1$ and $T_2$ into the cluster operator and the computational complexity is $O(N^6)$.

The multi-configuration method improves upon HF by considering not only the ground state but also the excited configurations like the CISD which includes single and double electron excitations from the reference Slater determinant. The wave function in CISD, $\Psi_{CISD}$, can be written as:

$$|\Psi_{CISD}\rangle = c_0|\Psi_{HF}\rangle + \sum_{ia} c_i^a |\Psi_i^a\rangle + \sum_{ijab} c_{ij}^{ab}|\Psi_{ij}^{ab}\rangle$$

where $\Psi_i^a$ and $\Psi_{ij}^{ab}$ represent singly and doubly excited determinants, respectively.

The $c_0$, $c_i^a$, $c_{ij}^{ab}$ are coefficients for the reference and excited determinants. Its computational complexity scales polynomially with the basis functions.

The multi-reference configuration interaction method builds upon the concept of employing multiple reference states to describe a system's electronic structure more comprehensively. It starts with a set of reference configurations, usually determined by a Complete Active Space Self-Consistent Field (CASSCF) calculation, which provides a balanced description of the molecule's ground and excited electronic states within a specified active space. The total wavefunction in MRCI, $\Psi_{MRCI}$, is expressed as an expansion over the reference configurations ($\Psi_i$) and additional configurations ($\Psi_a$) generated by exciting electrons from the occupied orbitals in the reference configurations to the virtual orbitals:

$$|\Psi_{MRCI}\rangle = \sum_i c_i |\Psi_i\rangle + \sum_a c_a |\Psi_a\rangle$$

Here, $c_i$ and $c_a$ are the coefficients for the reference and excited configurations, respectively, optimized to minimize the total energy of the system under the variational principle. The computational complexity of MRCI is one of its most significant challenges. The number of excited configurations ($\Psi_i$) grows combinatorially with the size of the active space and the level of excitations considered (singles, doubles, triples, etc.). This combinatorial explosion means the cost of MRCI calculations can quickly become prohibitive as the system size increases or as more extensive correlation effects are included.

The FCI method provides an exact solution to the electronic Schrödinger equation within the basis set limit by considering all possible excitations from the reference determinant. The FCI wave function, $\Psi_{FCI}$ can be expressed as:

$$|\Psi_{FCI}\rangle = c_0|\Psi_{HF}\rangle + \sum_{ia} c_i^a |\Psi_i^a\rangle + \sum_{ijab} c_{ij}^{ab}|\Psi_{ij}^{ab}\rangle + \cdots + \sum_{all\ excitations} c|\Psi_{excited}\rangle$$

The FCI method encompasses all singly, doubly, triply, and higher excited configurations, making it the most accurate but computationally intensive method. The computational complexity of FCI scales factorially with the number of basis functions, which restricts its application to very small systems. Despite its computational demand, FCI's accuracy is unparalleled, serving as a benchmark for other methods.

Density Matrix Renormalization Group (3,4) is another powerful numerical algorithm initially developed for low-dimensional quantum systems within condensed matter physics but has since been adapted to quantum chemistry. The essence of DMRG lies in the construction of a system block and an environment block, which are gradually built up by adding sites (orbitals) to the system. The wave function in DMRG is represented as a Matrix Product State (MPS):

$$|\Psi_{MPS}\rangle = \sum_{\{n\}} A^{n_1} A^{n_2} \dots A^{n_N} |n_1 n_2 \dots n_N\rangle$$

where N is the number of orbitals, $n_i$ denotes the occupancy of the ith orbital, $A^{n_i}$ and are matrices associated with each state of the orbital. The key is to find the optimal set of matrices that minimize the energy for a given Hamiltonian. The computational complexity of DMRG depends primarily on the bond dimension (m) of the MPS, scaling as $O(m^3)$ for the simplest cases but can increase depending on the system's specifics. The choice of m balances accuracy and computational cost.

Quantum Monte Carlo method is an advanced computational technique used in quantum chemistry (5,6). The typical one is Variational Monte Carlo (VMC) which is a stochastic approach that uses Monte Carlo integration to evaluate the expectation values of observables for quantum systems. It is particularly useful for studying systems where the wave function can be reasonably approximated by a parameterized ansatz. In VMC, the energy of a system is computed as:

$$E(\theta) = \frac{\langle \Psi(\theta)|H|\Psi(\theta)\rangle}{\langle \Psi(\theta)|\Psi(\theta)\rangle} = \frac{\sum_{x,x'}\langle \Psi(\theta)|x\rangle\langle x|H|x'\rangle\langle x'|\Psi(\theta)\rangle}{\sum_{x''}\langle \Psi(\theta)|x''\rangle\langle x''|\Psi(\theta)\rangle}$$

$$= \frac{\sum_x \left( \left( \sum_{x'} \langle x|H|x'\rangle\langle x'|\Psi(\theta)\rangle \middle/ \langle x|\Psi(\theta)\rangle \right) \langle \Psi(\theta)|x\rangle\langle x|\Psi(\theta)\rangle \right)}{\sum_{x''}\langle \Psi(\theta)|x''\rangle\langle x''|\Psi(\theta)\rangle}$$

where $|\Psi(\theta)\rangle$ is the trial wave function parameterized by a set of variational parameters $\theta$, and $H$ is the Hamiltonian of the system. x(x', x") are the computation basis and $\langle \Psi(\theta)|x\rangle\langle x|\Psi(\theta)\rangle$ are the probability amplitude of the wave function ansatz. The variational principle ensures that $E(\theta)$ is an upper bound to the ground state energy, which can be minimized by adjusting variational parameters. The computational complexity of VMC is influenced by the number of variational parameters and the complexity of evaluating the trial wave function and its derivatives.

## 1.2 Limitations of Quantum Algorithms for NISQ Computers

Enter the era of Noisy Intermediate-Scale Quantum (NISQ) computers, which promise a paradigm shift in computational quantum chemistry (7) . Despite their potential, quantum algorithms tailored for NISQ computers are not without limitations. These include issues related to coherence times, gate fidelities, and the overall quantum resource requirements, which currently restrict the practical applications of quantum computing in solving real-world chemical problems.

Quantum phase estimation is a powerful quantum algorithm for electronic structure calculations (8,9). However, it requires a deep quantum circuit and thus a long coherence time. A more practical way to perform quantum computation on NISQ devices is using a hybrid quantum-classical algorithm (10). Variational quantum eigensolver (VQE)(11) is such an algorithm to estimate the ground-state energy of a given Hamiltonian, based on a wavefunction ansatz or, more specifically, a parameterized quantum circuit. A parameterized wavefunction on a quantum computer

can usually be written as

$$|\Psi\rangle = U(\Theta)|\Psi_0\rangle = \prod_{i=1}^{N} U_i(\theta_i)|\Psi_0\rangle$$

$U_i(\theta_i)$ can either be associated with a specific electron excitation or correspond to a set of quantum gates depending on whether a physically motivated or a hardware heuristic ansatz is adopted. In both cases, the ground-state energy can be estimated by variationally minimizing the expected value of the Hamiltonian.

The mainstream quantum wave function ansatz, including physically-inspired approaches like the unitary coupled cluster (UCC) (12) and hardware-efficient ansatz (HEA) (13). Hardware Adaptable Ansatz (HAA) (14), along with their modified versions (15,16), are designed to capture the intricate many-body interactions between electrons. The UCC ansatz, such as UCCSD that relies on electron single and double excitation operators, offers a manageable number of variational parameters for training, but can lead to deeper circuit depths, necessitating certain simplifications. In contrast, the HEA approach offers relatively shallow circuit depths but requires a larger number of variational parameters for optimization.

### 1.3 Neural Network Quantum States (NNQS) Architecture

In response to these challenges, the neural network quantum states architecture emerges as a groundbreaking solution (17). Our QiankunNet leverages the power of machine learning to represent complex quantum wave functions, offering a novel approach to quantum chemistry that combines the computational advantages of classical neural networks with the rich, expressive capabilities of quantum systems (18). The architecture of an NNQS typically involves a neural network model that takes a quantum state's configuration (e.g., the spin orbital occupation number) as input and outputs a complex number that represents the amplitude of the wave function for that configuration.

### 1.4 QiankunNet-VQE Flowchart Overview

A pivotal development within this domain is the introduction of QiankunNet-VQE, a specific implementation of NNQS. The QiankunNet-VQE flowchart delineates a structured process wherein quantum states are efficiently parameterized and optimized through neural networks, facilitating the accurate simulation of quantum systems with reduced computational overhead.

The complete algorithm, which integrates the quantum computer with the classical transformer model, is illustrated in Figure 1. We begin with the quantum component, encompassing the selection of active orbitals, ansatz design, circuit optimization, and state extraction. In the classical component, our primary focus is on the training of our transformer model, QiankunNet (18,20). This process encompasses both the pre-training and formal optimization stages. This intricate process ensures that the quantum computer can effectively simulate the desired quantum system, laying the foundation for accurate predictions and insights. In representing wavefunctions using the NNQS approach, we employ four transformer decoder layers with an embedding size of $d_{model}$=32 and $n_{heads}$=4 for the amplitude part. For the phase part, we utilize three dense layers in the MLP with dimensions $N×512×512×4$, where $N$ is the number of spin orbitals. The total number of parameters is approximately $8.5×10^5$.

**1.4.1 Selection of active orbitals**. *In practice, the orbital selection step is not strictly necessary and is mainly dictated by the available quantum resources, particularly the number of qubits.* Given the current noisy intermediate-scale quantum (NISQ) computers, the selection of active orbitals serves to strike a balance between the practicality of quantum computation and the desired precision of the results (21). From a quantum chemistry perspective, determining the ground state energy of a molecule or the reaction barrier for a chemical reaction often involves electron correlation from only a few active orbitals. Numerous automated methods for selecting the necessary orbitals exist, including autoCAS (22), which relies on orbital entropy, and AVAS (23), which relies on projection onto atomic orbitals. These methods are all compatible with our current algorithms. For the purposes of this article, we have opted to freeze the lowest molecular orbitals based on their symmetry and orbital energy.

**1.4.2 Ansatz design.** In this article, we have evaluated both the UCCSD, hardware efficient ansatz (HEA) and hardware adaptable ansatz (HAA) and observed that they yield promising initial states for the transformer model, enhancing the accuracy of the final calculations.

**1.4.3 Circuit optimization.** This section details the classical component of the VQE experiment. The gradient-based optimization algorithms, such as steepest descent and Adam (24), are employed to iteratively update the variational parameters in order to minimize the energy of the system. To obtain the gradient of the energy with respect to the parameters, we utilize the parameter-shift rule (25). However, when dealing with a large number of parameters, as is often the case with the HEA approach, this can impose a significant experimental burden. To address this challenge, we randomly divide the parameters into groups and optimize them sequentially. We have observed that this strategy not only reduces the experimental burden but also helps avoid local minima, thus enabling a more efficient and robust optimization process.

**1.4.4 State extraction.** After the VQE has converged, various quantum tomography methods can be utilized to reconstruct the target state (26). When considering full configuration interaction, the number of possible configurations or determinants for a system with N spatial orbitals and N electrons is given by $(C_M^N)^2$, which rapidly increases with the size of the system. While quantum computers can efficiently store these configurations, measuring all of them is unrealistic and often unnecessary. It has been observed that important configurations tend to be sparse, particularly for ground states. In our algorithm, we attempted to include all possible configurations for smaller systems (typically containing fewer than 10 active orbitals) and employed Monte Carlo sampling for larger systems. The Monte Carlo sampling process follows these steps:

Initially, a configuration that aligns with the desired particle and spin number is randomly selected. By conducting projection measurements in quantum experiments, the weight associated with this configuration is determined. Subsequently, the weight is utilized to build the wave function and calculate the corresponding energy. The generated energy is then compared with the energy of the HF state. If the energy is lower, the configurations are accepted. If not, the corresponding probability is computed and compared with a newly generated random number. If the probability surpasses the random number, the configurations are accepted; otherwise, they are discarded. This iterative process continues until the energy variations introduced by new configurations are minimal, indicating that the sampling process has reached completion.

Pre-training aims to initialize the model parameters in a meaningful way, while formal optimization optimizes these parameters to improve the model's performance. This iterative process aims to refine the model's representation capabilities and enhance its predictive accuracy.

**1.4.5 Pretraining of QiankunNet with VQE.** The pretraining of large neural network models, such as Transformers, involves utilizing large datasets to train the model from end-to-end, typically utilizing unsupervised learning techniques. This process typically comprises two main stages: pretraining and fine-tuning. During pretraining in our case, the model is exposed to a vast collection of unlabeled data to acquire general language patterns and representations. The objective of pretraining is to initialize the model with meaningful representations such as the coefficients of configuration from VQE that can be fine-tuned for specific tasks like ground state energy. In our algorithms, we pretrain a Transformer model using Masked Language Modeling (MLM) and Next-Sentence Prediction (NSP) tasks. MLM helps the model learn relationships within quantum states by predicting masked tokens, while NSP enables it to understand sequence structures. For our QiankunNet-VQE model, we use configurations and coefficients from VQE experiments with a KL divergence loss function to refine the predicted wave functions. After pretraining, the model is fine-tuned on specific tasks such as quantum state classification and chemical reaction simulation, enhancing its performance and adaptability.

**1.4.6 Optimization of QiankunNet.** The Transformer, a formidable neural network model, has the capacity to learn molecular wavefunctions by representing the quantum state of a molecule as a sequence of tokens. It is trained using a blend of supervised and self-supervised learning techniques to predict the electronic structures of molecules. During training, the Transformer is presented with a sequence of atomic embeddings, which encode the quantum state of each atom in the molecule. The model then strives to predict the electronic properties of the molecule, including energy, dipole moment, and electron density. The Transformer architecture comprises multiple layers of self-attention and feedforward neural networks. The self-attention mechanism empowers the model to capture long-range interactions between atoms, while the feedforward networks introduce nonlinearity and depth to the model. The Transformer is trained using a combination of supervised learning and denoising autoencoding. In supervised learning, the model is trained to predict the electronic properties of molecules using labeled data containing quantum mechanical calculations. Denoising autoencoding is employed for pretraining by reconstructing corrupted versions of the input atomic embeddings. By harnessing the power of self-attention and feedforward networks, the Transformer can effectively learn the intricate quantum mechanical interactions between atoms in a molecule and represent them as a sequence of tokens. This capability facilitates accurate predictions of electronic properties and lays the foundation for further quantum chemical applications.

**1.5 Advantages of QiankunNet-VQE**

The advantages of QiankunNet-VQE are manifold. By harnessing the computational efficiency of neural networks, QiankunNet-VQE not only addresses the scalability issues faced by classical quantum chemistry methods but also circumvents the limitations inherent to NISQ computers. This innovative approach opens new avenues for simulating and understanding complex quantum systems, potentially revolutionizing the field of quantum chemistry with its ability to provide precise, scalable solutions for a wide range of chemical phenomena.

Utilizing quantum computers for quantum chemistry challenges presents a notable advantage over traditional methods, primarily their efficiency in simulating correlations or entanglements across multiple orbitals with straightforward quantum gate operations. Consider a four-orbital, four-electron system in the HF state, where electrons fill the four lowest energy states. The coupled cluster method, used to approximate the system's accurate wavefunction, shows that the computational complexity for including up to the highest, fourth-order excitations scales to the fourth power of the number of electrons and orbitals. Mapping these quadruple excitations onto a quantum computer typically necessitates deep circuits. Yet, achieving fourth-order excitations alone requires only four CNOT gates to facilitate the transition from the HF state to a configuration with quadruple excitations. A few Controlled-Z (CZ) gates can also induce interactions among various orbitals, producing the relevant configurations with shallow circuits. VQE experiments serve dual purposes: generating entanglement among multiple orbitals to ensure a wide range of configurations contribute to the energy, and optimizing the energy to its lowest possible value. Although the precision of the coefficients might be limited by the depth of the quantum circuits, the configurations and coefficients derived from these shallow circuits provide a solid foundation for further optimization in advanced models, including neural network-based quantum models. This approach leverages the entanglement and low-energy states produced by VQE to enhance the performance of more sophisticated quantum algorithms.

## 2. Configuration Analysis from Quantum Experiments.

### 2.1 Quantum Experimental circuits in Ref. 27

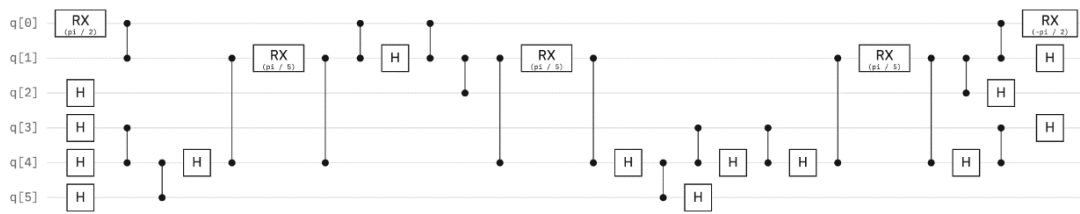

**Figure S1** The modified UCCSD quantum circuit for LiH molecule at quantum experiment with RX, CZ and H gates.

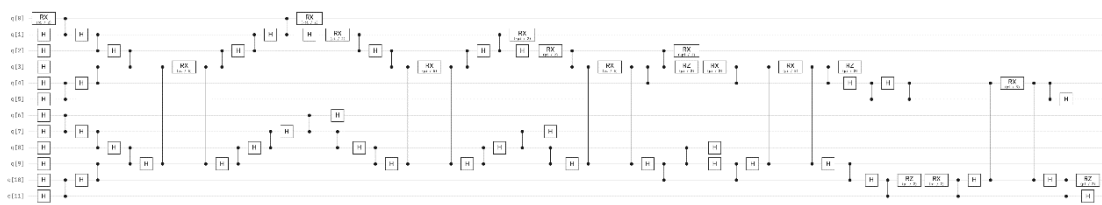

**Figure S2** The modified UCCSD quantum circuit for $F_2$ molecule at quantum experiment.

### 2.2 Lithium Hydride (LiH) System

**Table S1** The resulted configurations and corresponding coefficients from VQE experiment (VQE-Exp.), CISD, FCI methods for LiH at bond length 2.6 Å. The configurations taken directly from the experiment are marked in yellow.

|  | Selected configurations | Configuration coefficient |
|---|---|---|
| VQE-Exp. CAS(3o,2e) | 1 1 1 1 1 0 0 0 0 0 0 0 | 0.95376325+0j |
|  | 1 1 0 0 0 0 1 1 0 0 0 0 | -0.2073140+0j |
|  | 1 1 0 0 1 1 0 0 0 0 0 0 | -0.1482393+0j |
|  | 1 1 0 0 0 1 1 0 0 0 0 0 | 0.1252200+0j |
|  | 1 1 0 0 1 0 0 1 0 0 0 0 | 0.0870896+0j |
| CISD CAS(6o,4e) | 1 1 1 1 0 0 0 0 0 0 0 0 | 0.91549980+0j |
|  | 1 1 0 0 0 0 0 0 0 0 1 1 | -0.19606494+0j |
|  | 1 1 0 0 0 1 0 0 0 0 1 0 | -0.18091654+0j |
| | 1 1 0 0 1 0 0 0 0 0 0 1 | -0.18091654+0j |
| | 1 1 0 0 1 1 0 0 0 0 0 0 | -0.16517012+0j |
| FCI | 1 1 1 1 0 0 0 0 0 0 0 0 | 0.91525544+0j |
| | 1 1 0 0 0 0 0 0 0 0 1 1 | -0.19597694+0j |
| | 1 1 0 0 0 1 0 0 0 0 1 0 | -0.18110095+0j |
| | 1 1 0 0 1 0 0 0 0 0 0 1 | -0.18110095+0j |
| | 1 1 0 0 1 1 0 0 0 0 0 0 | -0.16561411+0j |

**Table S2** The energy from VQE-Exp., CISD, FCI methods for LiH at bond length 2.6 Å

| Method | Energy (Ha) |
|---|---|
| HF | -7.75840439 |
| VQE-Exp. | -7.80670244 |
| Exact Diag | -7.81362774 |
| CISD | -7.81734514 |
| FCI | -7.81739992 |

**Table S3** The resulted configurations and corresponding coefficients from VQE-Exp., CISD, FCI methods for LiH at bond length 3.0 Å. The configurations taken directly from the experiment are marked in yellow.

| | Selected configurations | Configuration coefficient |
|---|---|---|
| VQE-Exp. CAS(3o,2e) | 1 1 1 1 0 0 0 0 0 0 0 0 | 0.94108819+0j |
| | 1 1 0 0 0 0 1 1 0 0 0 0 | -0.20718356+0j |
| | 1 1 0 0 1 0 0 1 0 0 0 0 | 0.15827049+0j |
| | 1 1 0 0 0 1 1 0 0 0 0 0 | 0.14984020+0j |
| | 1 1 0 0 1 1 0 0 0 0 0 0 | -0.14310344+0j |
| CISD CAS(6o,4e) | 1 1 1 1 0 0 0 0 0 0 0 0 | 0.84875877+0j |
| | 1 1 0 0 1 1 0 0 0 0 0 0 | -0.29026586+0j |
| | 1 1 0 0 1 0 0 0 0 0 0 1 | -0.23662591+0j |
| | 1 1 0 0 0 1 0 0 0 0 1 0 | -0.23662591+0j |
| | 1 1 1 0 0 1 0 0 0 0 0 0 | 0.15525683+0j |
| FCI | 1 1 1 1 0 0 0 0 0 0 0 0 | 0.84833978+0j |
| | 1 1 0 0 1 1 0 0 0 0 0 0 | -0.29089478+0j |
| | 1 1 0 0 1 0 0 0 0 0 0 1 | -0.23674652+0j |
| | 1 1 0 0 0 1 0 0 0 0 1 0 | -0.23674652+0j |
| | 1 1 0 0 0 0 0 0 0 0 1 1 | -0.17642157+0j |

**Table S4** The energy from VQE-Exp., CISD, FCI methods for LiH at bond length 3.0 Å

| Method | Energy (Ha) |
|---|---|
| HF | -7.7108299 |
| VQE-Exp. | -7.78503864 |
| Exact Diag | -7.79841081 |
| CISD | -7.79875332 |
| FCI | -7.79884316 |

## 2.3 Fluorine Molecule (F$_2$) System

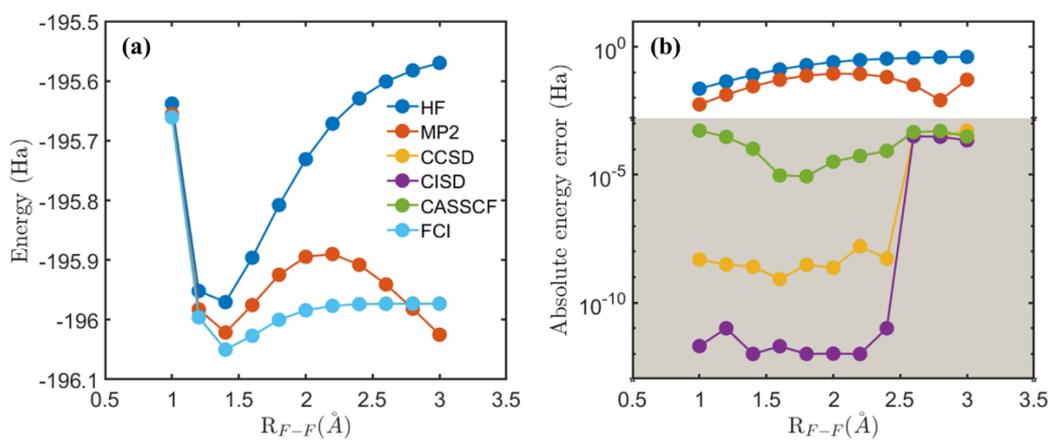

**Figure S3** The potential energy surface (a) and the absolute energy error (b) of $F_2$ calculated by different classical quantum chemistry methods like HF, MP2, CCSD, CISD, CASSCF, FCI.

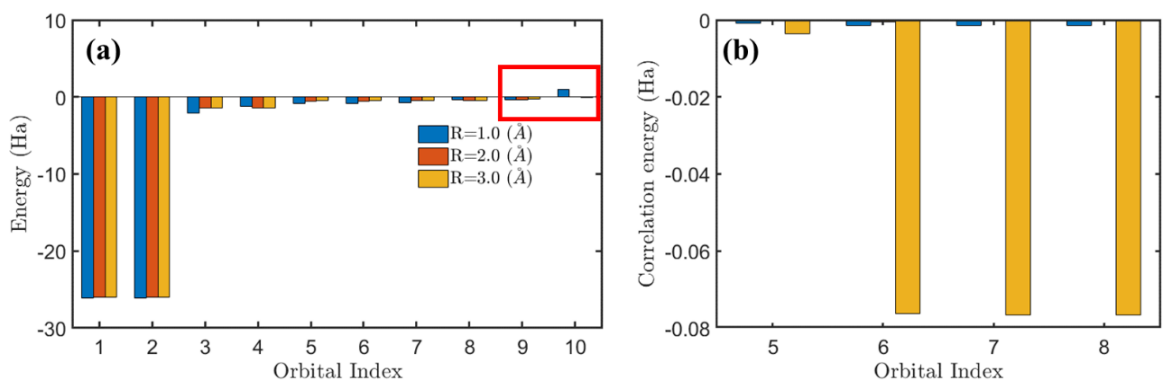

**Figure S4** (a) The molecular orbital energy and (b) The orbital correlation energy of $F_2$ at bond length of 1.0, 2.0 3.0 Å. The orbital correlation energy defines as $\Delta\epsilon_i = \epsilon_i - \epsilon_{ref}$, which $\epsilon_i$ is complete active space configuration interaction energy and $\epsilon_{ref}$ is the HF energy[21].

**Table S5** The energy from HF, CASSCF and FCI methods for LiH at different bond length.

| F₂ system basis:sto-3g, n_orb=10, n_elec=18 ||||| 
|---|---|---|---|---|
| Bond-length | HF | CASSCF(2o,2e) | FCI | diff |
| 1 | -195.6380415 | -195.655897 | -195.6610863 | 0.005189262 |
| 2 | -195.7310982 | -195.9834209 | -195.984138 | 0.000717117 |
| 3 | -195.569903 | -195.5649395 | -195.9732325 | 0.408293016 |

**Table S6** The resulted configurations and corresponding coefficients from VQE-Exp., CISD, FCI methods for $F_2$ at bond length 1.0 Å. The configurations taken directly from the experiment are marked in yellow.

|  | Selected configurations | Configuration coefficient |
|---|---|---|
| VQE-Exp. CAS(6o,10e) | 1 1 1 1 1 1 1 1 1 1 1 1 1 1 1 1 1 0 0 | -0.99775362+0j |
|  | 1 1 1 1 1 1 1 1 1 1 1 1 1 1 0 0 1 1 1 1 | 0.04428186+0j |

|  | 1 1 1 1 1 1 1 1 1 1 1 1 0 0 1 1 1 1 1 1 | 0.04106867+0j |
|---|---|---|
|  | 1 1 1 1 1 1 1 1 1 1 1 1 1 1 1 1 0 0 1 1 | 0.02579647+0j |
|  | 1 1 1 1 1 1 1 1 0 0 1 1 1 1 1 1 1 1 1 1 | -0.01086589+0j |
| CISD CAS(10o,18e) | 1 1 1 1 1 1 1 1 1 1 1 1 1 1 1 1 1 1 0 0 | 0.99599520+0j |
|  | 1 1 1 1 1 1 1 1 1 1 1 1 0 0 1 1 1 1 1 1 | -0.06507797+0j |
|  | 1 1 1 1 1 1 1 1 1 1 1 1 1 1 1 1 0 0 1 1 | -0.02584956+0j |
|  | 1 1 1 1 1 1 1 1 1 1 1 1 1 1 0 0 1 1 1 1 | -0.02584956+0j |
|  | 1 1 1 1 0 1 1 1 1 1 1 1 1 0 1 1 1 1 1 1 | -0.02555259+0j |
| FCI | 1 1 1 1 1 1 1 1 1 1 1 1 1 1 1 1 1 1 0 0 | 0.99599519+0j |
|  | 1 1 1 1 1 1 1 1 1 1 1 1 0 0 1 1 1 1 1 1 | -0.06507776+0j |
|  | 1 1 1 1 1 1 1 1 1 1 1 1 1 1 1 1 0 0 1 1 | -0.02584984+0j |
|  | 1 1 1 1 1 1 1 1 1 1 1 1 1 1 0 0 1 1 1 1 | -0.02584984+0j |
|  | 1 1 1 1 0 1 1 1 1 1 1 1 1 0 1 1 1 1 1 1 | -0.02555265+0j |

**Table S7** The energy from VQE-Exp., CISD, FCI methods for $F_2$ at bond length 1.0 Å

| Method | Energy (Ha) |
|---|---|
| HF | -195.6380415 |
| VQE-Exp. | -195.6501992 |
| Exact Diag | -195.6521913 |
| CISD | -195.6610863 |
| FCI | -195.6610863 |

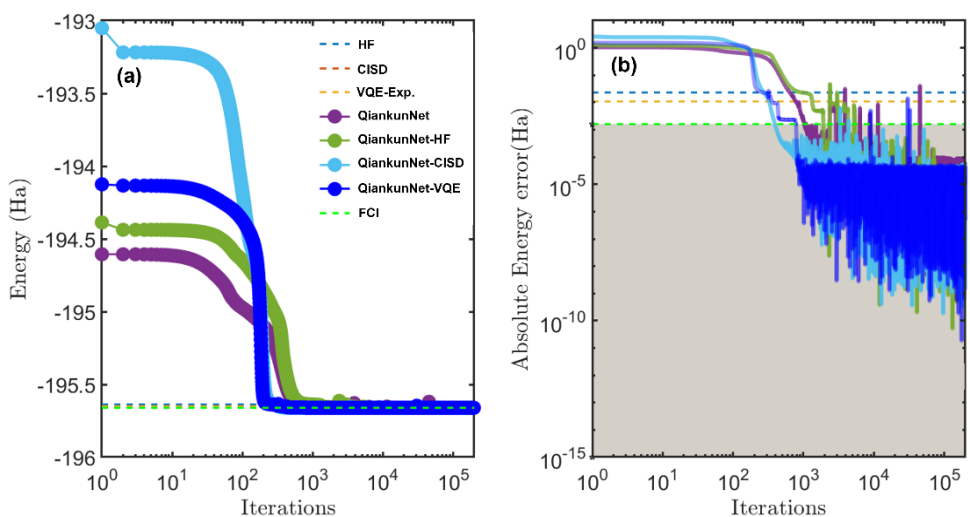

**Figure S5** (a) The energy and (b) absolute energy error converged process of $F_2$ with bond length of 1.0 Å for HF, CISD, VQE-Exp., QiankunNet, QiankunNet-HF, QiankunNet-CISD, QiankunNet-VQE and FCI.

**Table S8** The resulted configurations and corresponding coefficients from VQE-Exp., CISD, FCI methods for $F_2$ at bond length 2.0 Å. The configurations taken directly from the experiment are marked in yellow.

| | **Selected configurations** | **Configuration coefficient** |
|---|---|---|
| **VQE-Exp. CAS(6o,10e)** | 1 1 1 1 1 1 1 1 1 1 1 1 1 1 1 1 1 0 0 | -0.85787239+0j |
| | 1 1 1 1 1 1 1 1 1 1 1 1 1 1 1 1 0 0 1 1 | 0.51378104+0j |
| | 1 1 1 1 1 1 1 1 0 0 1 1 1 1 1 1 1 1 1 1 | 0.00548280+0j |
| | 1 1 1 1 1 1 1 1 1 1 1 1 1 1 0 0 1 1 1 1 | 0.00465198+0j |
| | 1 1 1 1 1 1 1 1 1 1 0 0 1 1 1 1 1 1 1 1 | 0.00242892+0j |
| **CISD CAS(10o,9e)** | 1 1 1 1 1 1 1 1 1 1 1 1 1 1 1 1 1 1 0 0 | 0.81654894+0j |
| | 1 1 1 1 1 1 1 1 1 1 1 1 1 1 1 1 0 0 1 1 | -0.57378388+0j |
| | 1 1 1 1 1 1 1 1 1 1 1 1 1 1 0 0 1 1 1 1 | -0.03261448+0j |
| | 1 1 1 1 1 1 1 1 1 1 1 1 0 0 1 1 1 1 1 1 | -0.03261448+0j |
| | 1 1 1 1 1 1 1 1 0 1 1 1 1 1 1 1 1 1 0 1 | -0.01874857+0j |
| FCI | **1 1 1 1 1 1 1 1 1 1 1 1 1 1 1 1 1 1 0 0** | **0.81654894+0j** |
| | **1 1 1 1 1 1 1 1 1 1 1 1 1 1 1 1 0 0 1 1** | **-0.57378388+0j** |
| | **1 1 1 1 1 1 1 1 1 1 1 1 1 1 0 0 1 1 1 1** | **-0.03261448+0j** |
| | **1 1 1 1 1 1 1 1 1 1 1 1 0 0 1 1 1 1 1 1** | **-0.03261448+0j** |
| | **1 1 1 1 1 1 1 0 1 1 1 1 1 1 1 1 1 1 0 1** | **-0.01874857+0j** |

**Table S9** The energy from VQE-Exp., CISD, FCI methods for $F_2$ at bond length 2.0 Å

| Method | Energy (Ha) |
|---|---|
| HF | -195.7310982 |
| VQE-Exp. | -195.9773569 |
| Exact Diag | -195.9966726 |
| CISD | -195.984138 |
| FCI | -195.984138 |

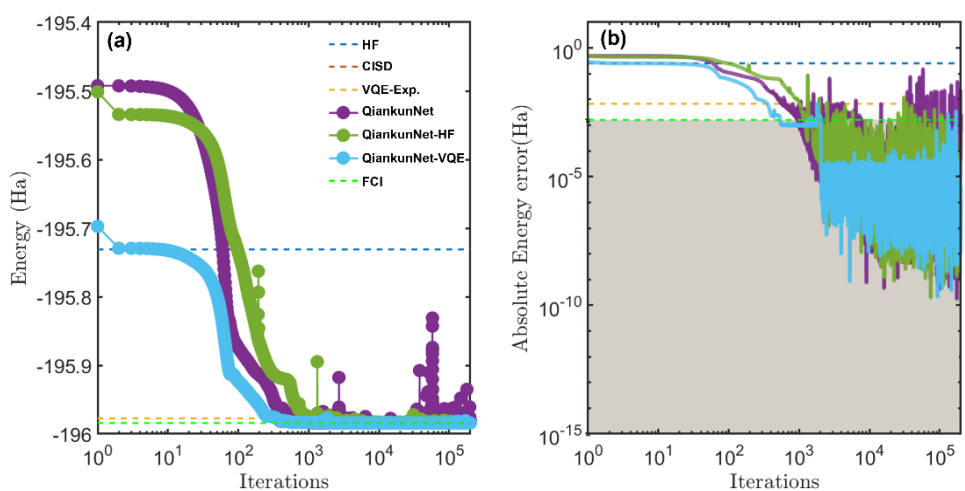

**Figure S6** (a) The energy and (b) absolute energy error converged process of $F_2$ with bond length of 2.0 Å for for HF, CISD, VQE-Exp., QiankunNet, QiankunNet-HF, QiankunNet-VQE and FCI.

**Table S10** The resulted configurations and corresponding coefficients from VQE-Exp.,

CISD, FCI methods for $F_2$ at bond length 3.0 Å. The configurations taken directly from the experiment are marked in yellow.

| | Selected configurations | Configuration coefficient |
|---|---|---|
| VQE-Exp. CAS(6o,4e) | 1 1 1 1 1 1 1 1 1 1 1 1 1 1 1 1 1 0 0 | -0.7404097+0j |
| | 1 1 1 1 1 1 1 1 1 1 1 1 1 1 1 1 0 0 1 1 | 0.67202901+0j |
| | 1 1 1 1 1 1 1 1 0 0 1 1 1 1 1 1 1 1 1 1 | -0.00635986+0j |
| | 1 1 1 1 1 1 1 1 1 1 0 0 1 1 1 1 1 1 1 1 | -0.00614850+0j |
| | 1 1 1 1 1 1 1 1 1 1 1 1 1 1 0 0 1 1 1 1 | -0.00316002+0j |
| CISD CAS(10o,18e) | 1 1 1 1 1 1 1 1 1 1 1 1 1 1 1 1 1 1 0 0 | 0.44158625+0j |
| | 1 1 1 1 1 1 1 1 1 1 1 1 1 1 1 1 0 0 1 1 | -0.43615982+0j |
| | 1 1 1 1 1 1 1 1 1 1 1 1 1 1 0 0 1 1 1 1 | -0.39233606+0j |
| | 1 1 1 1 1 1 1 1 1 1 1 1 0 0 1 1 1 1 1 1 | -0.39233606+0j |
| | 1 1 1 1 1 1 1 1 1 1 0 0 1 1 1 1 1 1 1 1 | 0.3917023+0j |
| FCI | 1 1 1 1 1 1 1 1 1 1 1 1 1 0 1 1 1 1 0 1 | 0.5020248+0j |
| | 1 1 1 1 1 1 1 1 1 1 1 1 0 1 1 1 1 1 1 0 | -0.5020248+0j |
| | 1 1 1 1 1 1 1 1 0 1 1 1 1 1 1 0 1 1 1 | -0.4964267+0j |
| | 1 1 1 1 1 1 1 1 0 1 1 1 1 1 1 1 1 0 1 1 | 0.4964267+0j |
| | 1 1 1 1 1 0 1 1 0 1 1 1 1 1 1 1 1 1 1 1 | -0.0043518+0j |

**Table S11** The energy from VQE-Exp., CISD, FCI methods for $F_2$ at bond length 3.0 Å

| Method | Energy (Ha) |
|---|---|
| HF | -195.569903 |
| VQE-Exp. | -195.9708747 |
| Exact Diag | -195.9729183 |

| CISD | -195.9730115 |
|---|---|
| FCI | -195.9732325 |

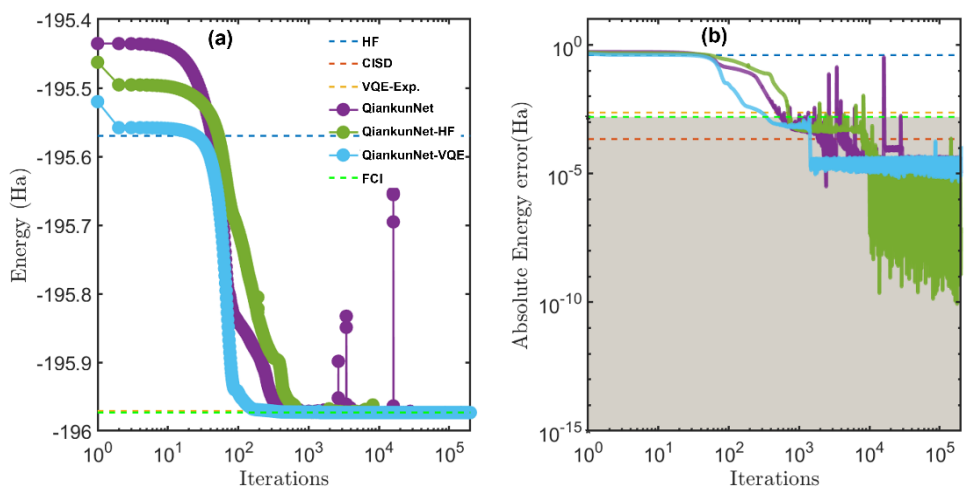

**Figure S7** (a) The energy and (b) absolute energy error converged process of $F_2$ with bond length of 3.0 Å for HF, CISD, VQE-Exp., QiankunNet, QiankunNet-HF, QiankunNet-VQE and FCI.

## 3. Simulations on Molecular Systems

### 3.1 $H_4$ Square System

**Table S12** The energy from UHF, MP2, CCSD, CISD, VQE simulation with CCSD ansatz (VQE-Simulator), FCI methods for $H_4$ at bond length 1.23 Å

| Method | Energy (Ha) |
|---|---|
| UHF | -1.9335630730 |
| MP2 | -1.9419070897 |
| CCSD | -1.9609443121 |
| CISD | -1.9593502913 |
| VQE-Simulator | -1.9675232627 |
| FCI | -1.9695121652 |

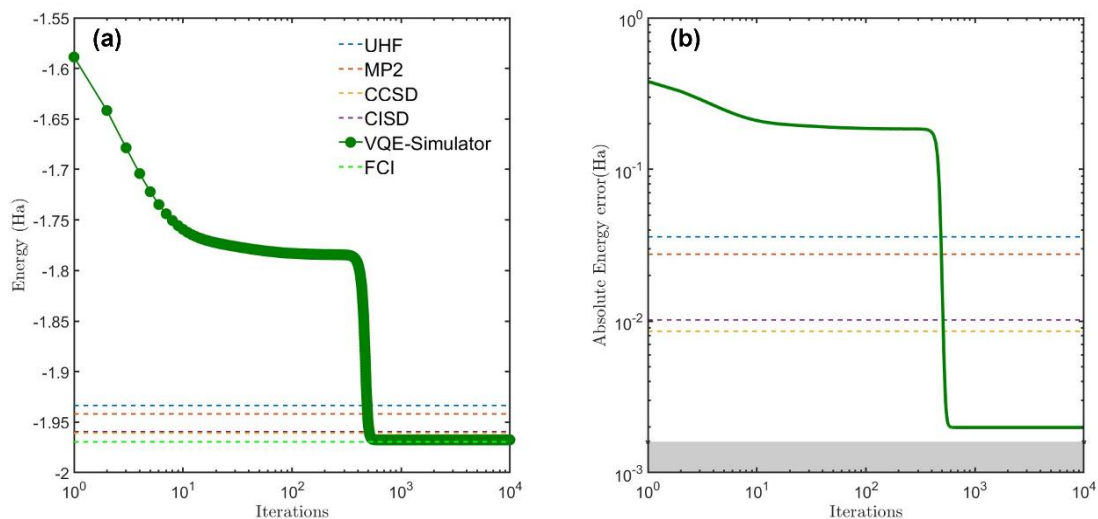

**Figure S8** The VQE simulation with UCCSD ansatz. (a) The energy and (b) absolute energy error converged process of $H_4$ with bond length of 1.23 Å for for HF, MP2, CCSD, CISD and FCI.

**Table S12** The converged variational parameters for $H_4$ system by using VQE-simulator with UCCSD ansatz

| Column1 | Column2 | Column3 | Column4 | Column5 | Column6 | Column7 | Column8 |
| --- | --- | --- | --- | --- | --- | --- | --- |
| -1.32786E-16 | -0.428663604 | 4.61034E-17 | -0.428663604 | -1.570783885 | 6.61731E-18 | -1.57078386 | 6.49073E-17 |
| 0.033910699 | 4.35828E-17 | -3.87066E-18 | 0.204108926 | 0.145098681 | 2.08384E-17 | 0.307425135 | -0.16259465 |
| -2.49479E-17 | -2.85121E-17 | -0.16259465 | 0.307425135 | 2.29319E-17 | 0.145098681 | 1.499434186 | -8.06982E-18 |
| 2.644E-17 | 0.039480117 | | | | | | |

## 3.2 Nitrogen ($N_2$) System

**Table S13** The minimum quantum resources required to achieve chemical accuracy include the number of qubits and the depth of the circuit for different wave functions ansatz including UCCSD, HEA and HAA, which are proposed to simulate the ground state energy of small molecular systems like $H_2$, $H_4$, $H_5^+$, $BeH_2$, $LiH$, $H_2O$, $N_2$.

| System | n_elec | Active Space (n_orb,n_elec) | Qubits | Circuit depth | | |
| --- | --- | --- | --- | --- | --- | --- |
| | | | | UCCSD | HEA | HAA |
| $H_2$ | 2 | (2,2) | 4 | 24 | 4 | 9 |
| H₄ | 4 | (4,4) | 8 | 339 | 20 | 33 |
| H₅⁺ | 4 | (5,4) | 10 | 772 | 72 | 48 |
| BeH₂ | 6 | (5,2) | 10 | 792 | 91 | 60 |
| LiH | 4 | (6,4) | 12 | 1415 | 95 | 84 |
| H₂O | 10 | (7,10) | 14 | 2286 | 107 | 93 |
| N₂ | 14 | (7,8) | 14 | 3308 | 123 | 93 |

Next, we test our algorithm in the nitrogen system potential energy surface with different classical algorithms.

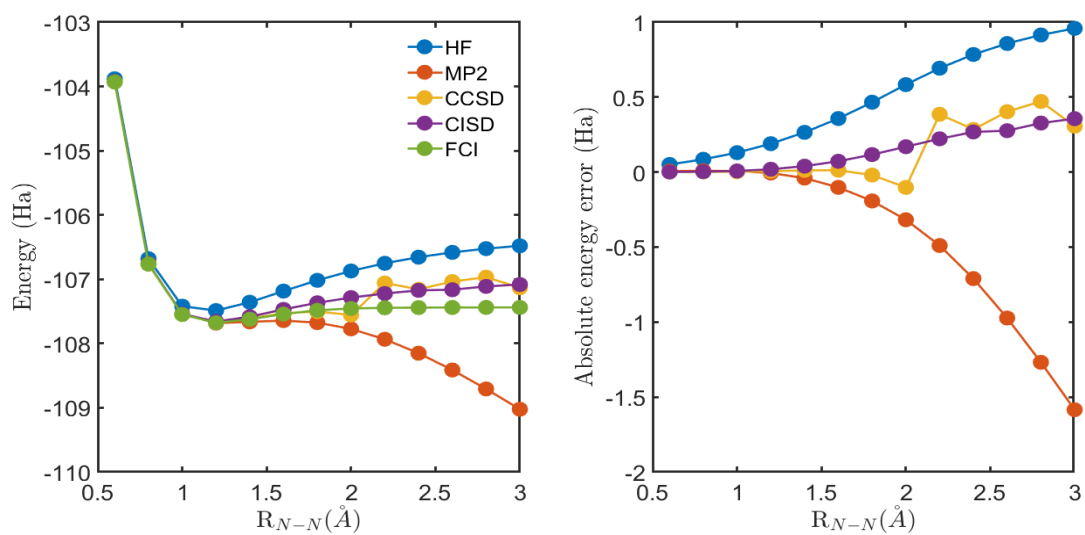

**Figure S9** The potential energy surface for $N_2$ molecular system in sto-3g basis.

**Table S14** The classical calculation results including Hartree-Fock method(HF), MP2, CCSD, CISD and FCI for the potential energy surface for $N_2$ molecular system in sto-3g basis.

| N₂ system basis:sto-3g, n_orb=10, n_elec=14 | | | | | |
|---|---|---|---|---|---|
| R | HF | MP2 | CCSD | CISD | FCI |
| 0.6 | -103.88071560 | -103.92458856 | -103.93126718 | -103.93078901 | -103.93173260 |
| 0.8 | -106.68080245 | -106.75726655 | -106.76517475 | -106.76330922 | -106.76630791 |
| 1.0 | -107.41953245 | -107.54208994 | -107.54672292 | -107.54138800 | -107.54930095 |

| 1.2 | -107.48778392 | -107.68317012 | -107.67143909 | -107.65830890 | -107.67733974 |
| 1.4 | -107.35781544 | -107.66309320 | -107.61164385 | -107.58327377 | -107.62317417 |
| 1.6 | -107.18484646 | -107.64375370 | -107.52935156 | -107.46955087 | -107.54208576 |
| 1.8 | -107.01732690 | -107.67563643 | -107.50298938 | -107.36645091 | -107.48345735 |
| 2.0 | -106.87150404 | -107.77279691 | -107.55698465 | -107.28567167 | -107.45515559 |
| 2.2 | -106.75183126 | -107.93423775 | -107.05903814 | -107.22273315 | -107.44485854 |
| 2.4 | -106.65660804 | -108.15127717 | -107.15652145 | -107.17337255 | -107.44130628 |
| 2.6 | -106.58195533 | -108.41274311 | -107.03653405 | -107.16346842 | -107.43978526 |
| 2.8 | -106.52411517 | -108.70717965 | -106.96764608 | -107.11158961 | -107.43885794 |
| 3.0 | -106.47984262 | -109.02326014 | -107.13237911 | -107.08194999 | -107.43846053 |

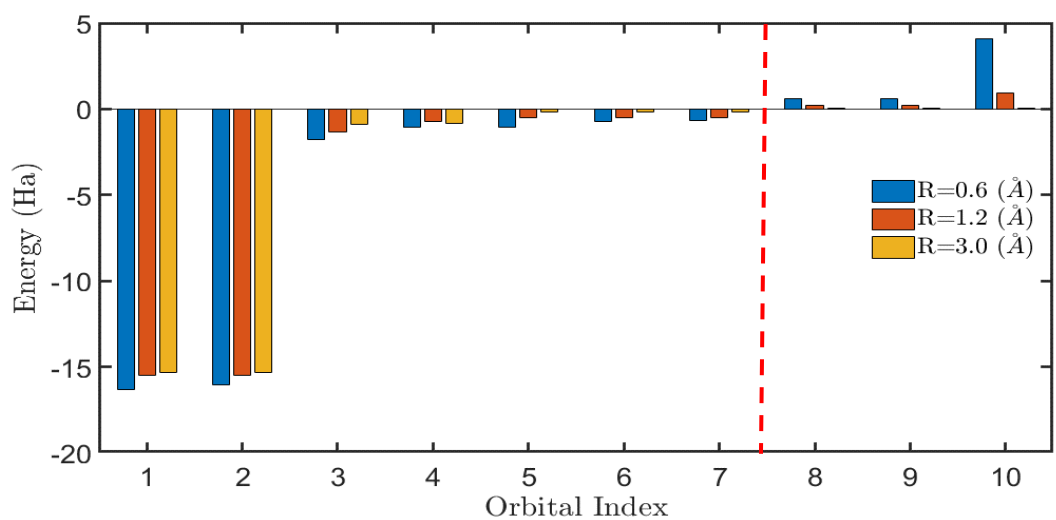

**Figure S10** The molecular orbital energy for $N_2$ molecular system at bond length of 0.6, 1.2 3.0 A.

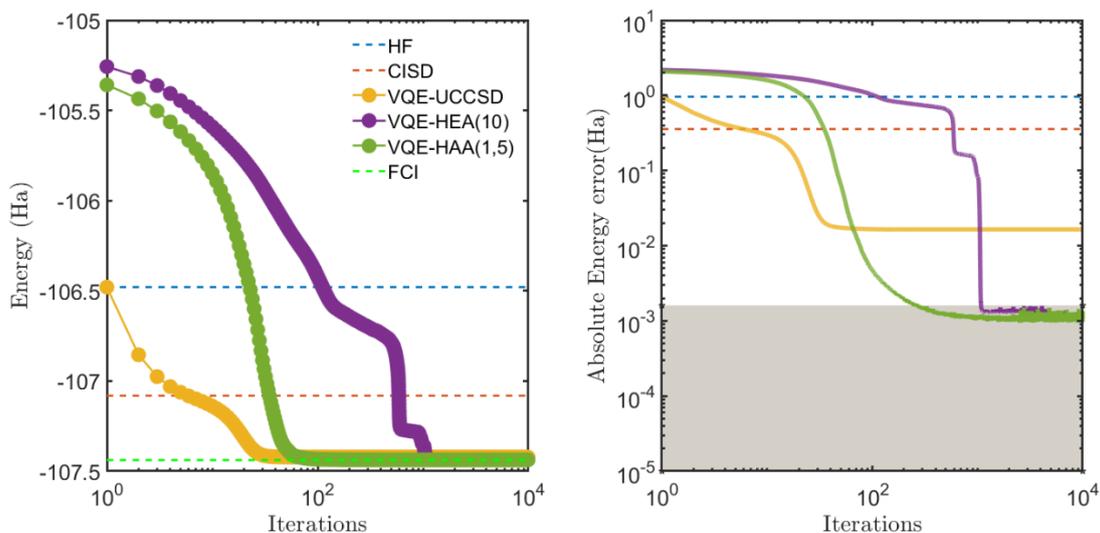

**Figure S11** (a) The energy and (b) absolute energy error converged process of $N_2$ with bond length of 3.0 A for VQE with UCCSD, HEA(10), HAA(1,5) ansatz.

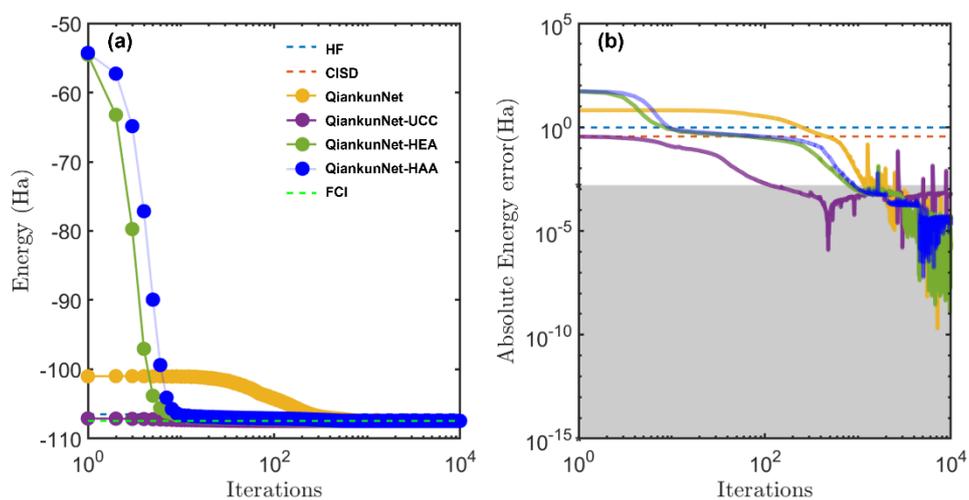

**Figure S12** (a) The energy and (b) absolute energy error converged process of $N_2$ with bond length of 3.0 A for HF, CISD and QiankunNet-VQE with different pre-training data like with UCCSD ansatz (QiankunNet-UCC), HEA ansatz (QiankunNet-HEA), HAA ansatz (QiankunNet-HAA).

**Table S14** The top five with the most weight configurations and corresponding coefficients from VQE simulation with UCCSD, HEA and HAA ansatz, CISD, FCI methods for $N_2$ at bond length 3.0 Å. The results from VQE simulation are marked in orange.

|  | Selected configurations | Configuration |
|---|---|---|

|  |  | coefficient |
|---|---|---|
| **VQE_UCC CAS(7o,8e)** | 1 1 1 1 1 1 1 1 1 1 1 1 1 1 0 0 0 0 0 0 | 0.391933568951+0j |
|  | 1 1 1 1 1 1 1 1 0 0 0 0 0 0 1 1 1 1 1 1 | -0.332345362851+0j |
|  | 1 1 1 1 1 1 1 1 1 1 0 0 0 0 1 1 1 1 0 0 | 0.256505652947+0j |
|  | 1 1 1 1 1 1 1 1 0 0 1 1 1 1 0 0 0 0 1 1 | -0.232421312942+0j |
|  | 1 1 1 1 1 1 1 1 0 0 0 0 1 1 1 1 0 0 1 1 | 0.162459884296+0j |
| **VQE_HEA(10) CAS(7o,8e)** | 1 1 1 1 1 1 1 1 0 1 0 1 0 1 0 1 0 1 0 1 | -0.99686855114 +0.07662705936j |
|  | 1 1 1 1 1 1 0 1 0 1 0 1 0 1 0 1 0 1 1 1 | 0.01938301272 -0.00149246605j |
|  | 1 1 1 1 1 1 1 1 0 0 1 0 1 0 1 1 1 0 1 1 | 0.00034765093 -0.00030606814j |
|  | 1 1 1 1 1 1 1 1 0 1 0 1 0 1 0 0 0 1 0 1 | -0.000227028526 +0.00039393030j |
|  | 1 1 1 1 1 1 1 1 0 1 0 1 0 0 1 0 1 0 1 0 | 0.000446011230 -6.830400018e-05j |
| **VQE_HAA(1,5) CAS(7o,8e)** | 1 1 1 1 1 1 1 1 1 0 1 0 1 0 1 0 1 0 1 0 | -0.94834379774 +0.31662645160j |
|  | 1 1 1 1 1 1 1 0 1 0 1 0 1 0 1 0 1 0 1 1 | 0.01868060122 -0.00623218486j |
|  | 1 1 1 1 1 1 0 0 1 0 1 0 1 0 1 0 1 0 1 0 | 0.00045673001 +0.00044144609j |
|  | 1 1 1 1 1 1 1 1 1 1 0 1 1 0 1 0 1 0 1 0 | 0.00026443597 +0.00047727070j |
|  | 1 1 1 1 1 1 1 1 1 1 1 1 1 0 1 0 1 0 1 0 | 0.000384411748 -0.000346797680j |
| **CISD CAS(10o,14e)** | 1 1 1 1 1 1 1 1 0 1 1 1 1 0 0 0 1 0 0 1 | -0.536177431005+0j |
|  | 1 1 1 1 1 1 1 1 0 1 1 0 1 0 0 0 1 1 0 | -0.328504341550+0j |
|  | 1 1 1 1 1 1 1 1 0 0 1 1 1 1 0 0 0 0 1 1 | 0.229502792347+0j |
|  | 1 1 1 1 1 1 1 1 0 1 1 1 0 0 0 0 1 0 1 | 0.216793211688+0j |
|  | 1 1 1 1 1 1 1 1 0 1 1 0 1 1 1 0 0 0 0 1 | 0.216271393001+0j |
| **FCI** | 1 1 1 1 1 1 1 1 1 1 1 0 0 1 1 0 0 1 0 0 | 0.2409553991242+0j |
|  | 1 1 1 1 1 1 1 1 1 1 0 1 1 0 0 1 1 0 0 0 | 0.2407608878579+0j |
|  | 1 1 1 1 1 1 1 1 1 1 1 1 1 1 0 0 0 0 0 0 | 0.2310060543337+0j |
|  | 1 1 1 1 1 1 1 1 0 1 1 0 1 1 1 0 0 0 0 1 | -0.228241392675+0j |
|  | 1 1 1 1 1 1 1 1 0 1 1 1 1 0 0 0 1 0 0 1 | 0.2282014864972+0j |

## 3.3 Hydrogen Chain ($H_{10}$) System

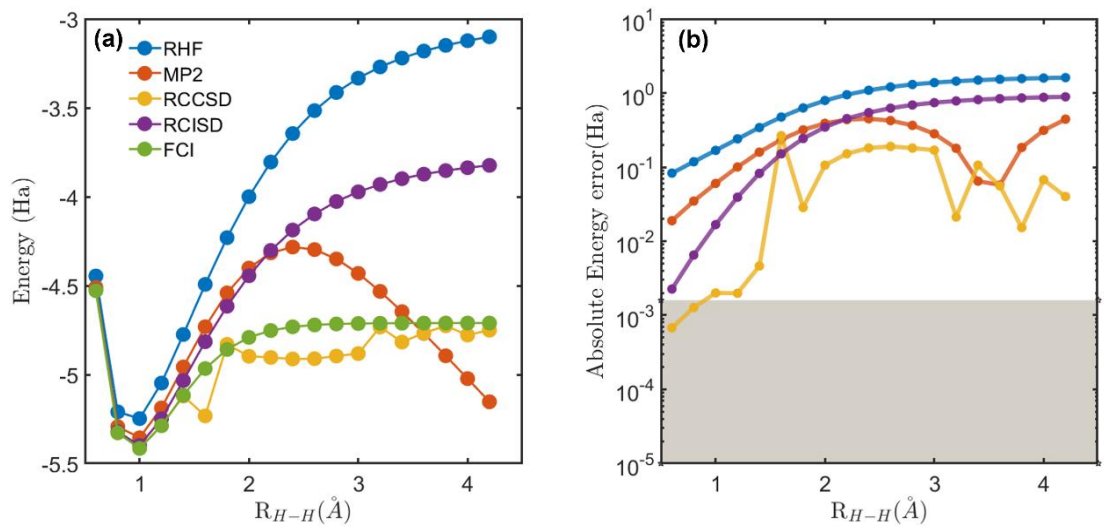

**Figure S13** (a) The energy and (b) absolute energy error converged process of $H_{10}$ potential energy surface for RHF, MP2, RCISD, RCCSD and FCI.

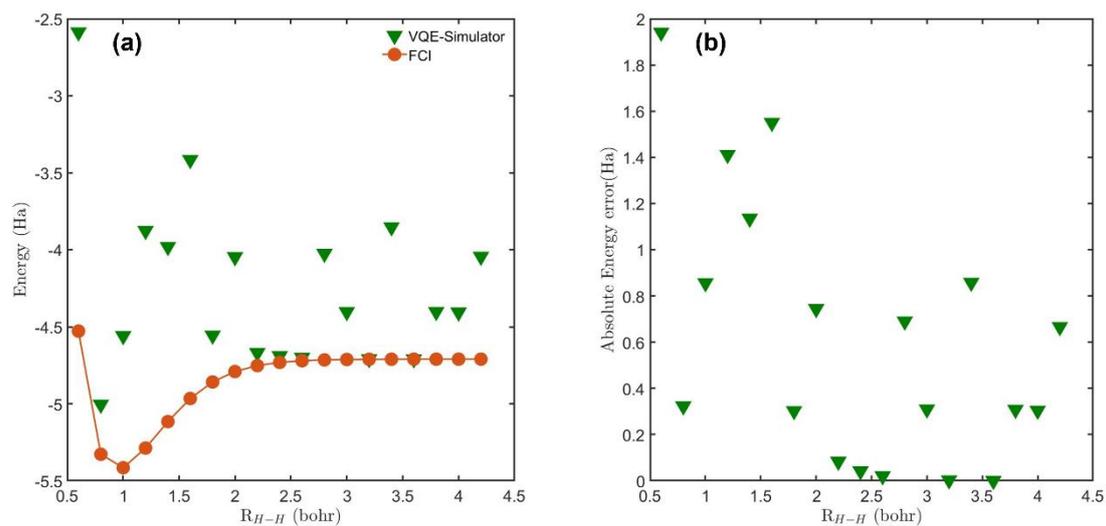

**Figure S14** VQE simulation results with UCCSD ansatz. (a) The energy and (b) absolute energy error converged process of $H_{10}$ potential energy surface.

**Table S15** The calculation results by UHF, UCISD, CCSD, CISD, QiankunNet, QiankunNet-VQE and FCI for the potential energy surface for the $H_{10}$ system in sto-6g basis.

| R(bohr) | R(Å) | UHF | UCISD | QiankunNet | QiankunNet-VQE | FCI |
|---|---|---|---|---|---|---|
| 1.0 | 0.529177 | -3.751737768 | -3.822900886 | -3.824385153 | -3.824385883 | -3.824385889 |
| 1.2 | 0.6350124 | -4.678053831 | -4.763622688 | -4.766375546 | -4.766376074 | -4.76637616 |
| 1.4 | 0.7408478 | -5.098618779 | -5.200263779 | -5.205086494 | -5.205093288 | -5.205093342 |
| 1.6 | 0.8466832 | -5.256281358 | -5.376186729 | -5.384321575 | -5.384360758 | -5.384360761 |
| 1.8 | 0.9525186 | -5.277448773 | -5.373758617 | -5.424365712 | -5.424385369 | -5.424385395 |
| 2.0 | 1.058354 | -5.231365453 | -5.340349582 | -5.389564310 | -5.389626111 | -5.389626114 |
| 2.2 | 1.1641894 | -5.153297935 | -5.268822907 | -5.316738408 | -5.316839261 | -5.316839354 |
| 2.4 | 1.2700248 | -5.065709686 | -5.179964713 | -5.227683836 | -5.22793288 | -5.227936932 |
| 2.6 | 1.3758602 | -4.982691147 | -5.086103888 | -5.136104375 | -5.136346015 | -5.136347009 |
| 2.8 | 1.4816956 | -4.91182638 | -4.994535507 | -5.049852571 | -5.050244988 | -5.050245415 |
| 3.0 | 1.587531 | -4.855611647 | -4.909741826 | -4.973608225 | -4.974243655 | -4.974243848 |
| 3.2 | 1.6933664 | -4.813234694 | -4.835291019 | -4.909668322 | -4.910382964 | -4.910383243 |
| 3.4 | 1.7992018 | -4.78235733 | -4.772409811 | -4.858190414 | -4.858866712 | -4.858866872 |
| 3.6 | 1.9050372 | -4.760346818 | -4.722438192 | -4.817867657 | -4.81870106 | -4.818701063 |
| 3.8 | 2.0108726 | -4.744874013 | -4.681899382 | -4.788090584 | -4.788255601 | -4.78825565 |
| 4.0 | 2.116708 | -4.734094083 | -4.648618823 | -4.764823913 | -4.765701079 | -4.765701332 |
| 4.2 | 2.2225434 | -4.726628874 | -4.616929121 | -4.748980476 | -4.749296933 | -4.749297019 |